\documentclass[letterpaper, 10 pt, conference]{IEEEtran} 

\usepackage{graphicx}
\usepackage{caption}
\graphicspath{{./}}
\usepackage[caption=false]{subfig}
\usepackage{placeins}
\usepackage{booktabs}
\usepackage[english]{babel}

\usepackage{soul, color}
\usepackage[hidelinks]{hyperref}

\IEEEoverridecommandlockouts     
 
\title{\LARGE \bf
Penobscot Dataset: Fostering Machine Learning Development for Seismic Interpretation
}

\author{Lais Baroni$^{1}$, Reinaldo Mozart Silva$^{1}$, Rodrigo S. Ferreira$^{1}$, Daniel Civitarese$^{1}$ \\ Daniela Szwarcman$^{1}$, Emilio Vital Brazil$^{1}$
\thanks{$^{1}$Reinaldo Mozart Silva, Lais Baroni, Rodrigo S. Ferreira, Daniel Civitarese, Daniela Szwarcman and Emilio Vital Brazil are with IBM Research,
        Av. Pasteur 138/146, 22290-240, Botafogo, Rio de Janeiro, Brazil
        {\tt\small \{rmozart, lbaroni, rosife, sallesd, daniszw, evital\} at br.ibm.com}}
}

\begin{document}

\maketitle

\thispagestyle{empty}
\pagestyle{empty}

\begin{abstract}

We have seen in the past years the flourishing of machine and deep learning algorithms in several applications such as image classification and segmentation, object detection and recognition, among many others. This was only possible, in part, because datasets like ImageNet -- with +14 million labeled images -- were created and made publicly available, providing researches with a common ground to compare their advances and extend the state-of-the-art.
Although we have seen an increasing interest in machine learning in geosciences as well, we will only be able to achieve a significant impact in our community if we collaborate to build such a common basis. This is even more difficult when it comes to the Oil \& Gas industry, in which confidentiality and commercial interests often hinder the sharing of datasets with others.
In this letter, we present the Penobscot interpretation dataset, our contribution to the development of machine learning in geosciences, more specifically in seismic interpretation. The Penobscot 3D seismic dataset was acquired in the Scotian shelf, offshore Nova Scotia, Canada. The data is publicly available and comprises pre- and pos-stack data, 5 horizons and well logs of 2 wells. However, for the dataset to be of practical use for our tasks, we had to reinterpret the seismic, generating 7 horizons separating different seismic facies intervals. The interpreted horizons were used to generated +100{,}000 labeled images for inlines and crosslines.
To demonstrate the utility of our dataset, results of two experiments are presented.

\end{abstract}

\section{INTRODUCTION}

The seismic reflection method is paramount for the location of possible hydrocarbon accumulation in the subsurface. Besides providing structural imaging and high area coverage in a short time, this geophysical method, in conjunction
with additional data, provides invaluable information about rock properties, fluid content and lithology.

In this context, using the knowledge provided by the seismic method, one may evaluate not only the location of possible reservoirs but also the economic viability with reasonable accuracy, therefore reducing risk and potential losses. However, the interpretation procedure of the seismic data is a human-intensive and time-consuming task, performed by geoscientists who are continually dealing with tight deadlines and the ever-increasing size of datasets \cite{randen2000}. 

In this scenario, researchers have proposed the application of computer-aided systems to assist geoscientists in several tasks involving seismic interpretation. For example, \cite{guillen2015, zhang2014, gao2011, mattos2017assessing} aim at the identification of specific structures on seismic images and \cite{west2002, song2017} propose techniques to automate part of the seismic facies analysis process. 

Other domains facing similar problems are using neural networks and machine/deep learning techniques with great success to support tasks that deal with high volumes of data and are considered human-centered, for instance, image classification and segmentation \cite{Badrinarayanan2015, Shelhamer}, and object detection and recognition \cite{Redmon_2016_CVPR, NIPS2015_5638}. Nevertheless, these methods require training datasets with a sufficient amount of data for the testing and validation of the proposed methodology. For the machine learning community, it is a common practice to make these datasets publicly available. Some examples are MNIST \cite{lecun2010} ($\sim$60k images), PASCAL-VOC \cite{everingham2010} ($\sim$40k images), MS-COCO \cite{lin2014} ($\sim$330k images), and ImageNet \cite{deng2009} ($\sim$14 million images). These datasets allow researchers to compare their advances, extend the state-of-the-art, and find new possible applications.

Although we have seen an increasing interest in machine learning in geosciences, to the best of our knowledge, there is no public labeled dataset targeting the seismic interpretation task. In this sense, we propose a new dataset to be publicly available.
The Penobscot interpretation dataset consists of 7 horizons and +100{,}000 labeled seismic images derived from the Penobscot 3D seismic data \cite{osr}, already in the public domain. The seismic lines were segmented in different portions based on their seismic facies. The proposed dataset has already been used in some applications, for example, the works of Chevitarese \textit{et al.} \cite{chevitarese2018deep, chevitarese2018seismic, chevitarese2018transfer}, which will be discussed in Section \ref{sec:experiments}.

The present paper is organized as follows: in the next section we describe the regional geology where the Penobscot survey lies. In Sections \ref{sec:seismic_data} and \ref{sec:seismic_interpretation} we present the Penobscot 3D seismic dataset and our interpretation procedure. Section \ref{sec:dataset} presents the proposed dataset and describes its main characteristics. Section \ref{sec:experiments} briefly discuss two works in which the proposed datasets was employed and we conclude our work in Section \ref{sec:conclusions}. 

\section{GEOLOGICAL SETTINGS}
\label{sec:geological_settings}

\begin{figure}
    \centering
    \includegraphics[width=0.5\textwidth]{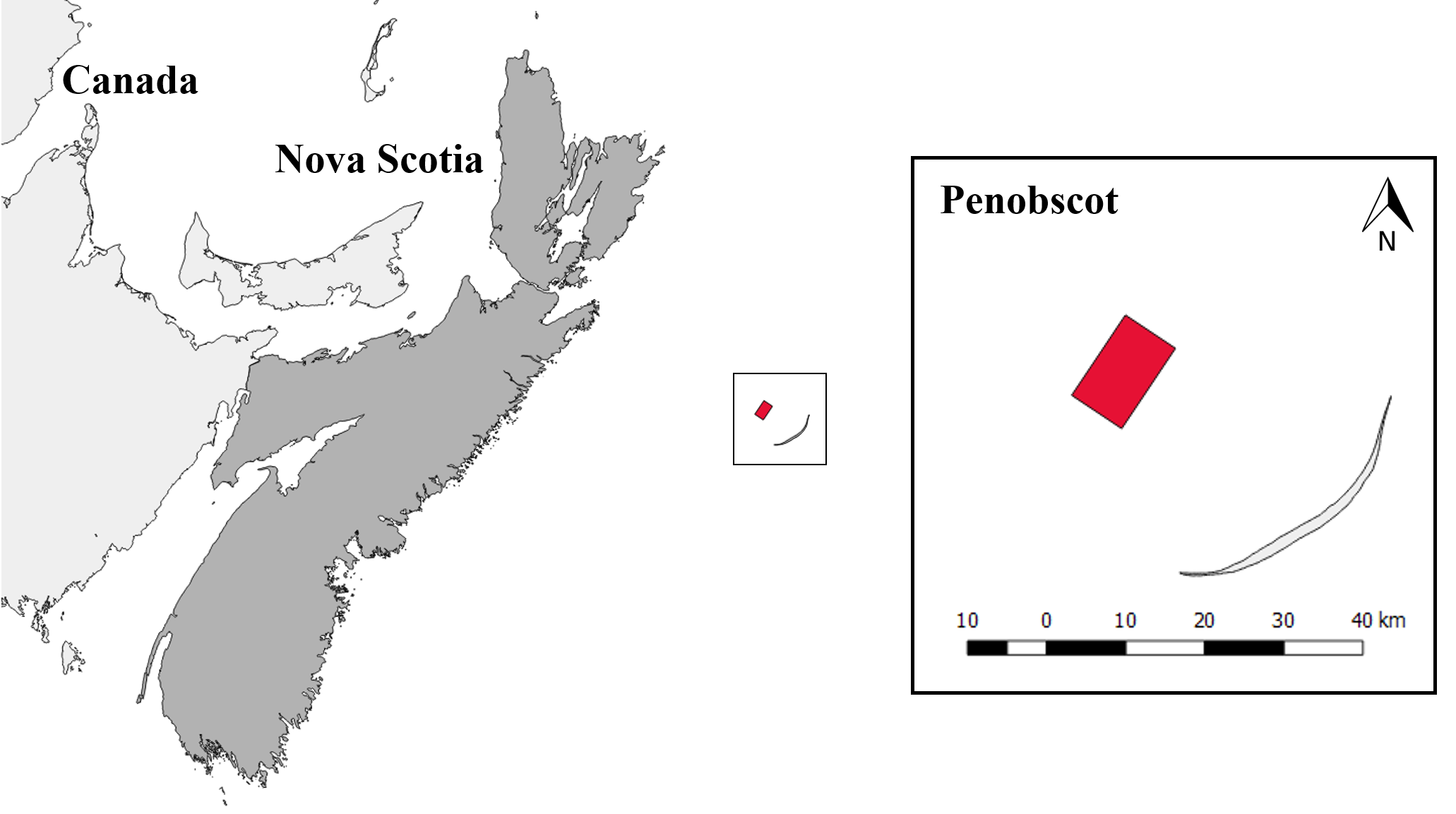}
    \caption{Location of the Penobscot 3D survey in the Scotian Basin, offshore Nova Scotia, Canada.}
    \label{fig:location} 
\end{figure}

The Penobscot 3D dataset \cite{osr} was acquired in the Scotian Basin, located on the Scotian Shelf, offshore Nova Scotia, Canada (\autoref{fig:location}). The basin was formed during the break-up of Pangaea -- separation of North American and African plates -- and covers an area of 300,000 km\textsuperscript{2} with sediment maximum thickness of 18 km \cite{hansen2004}. The rifting process that took place from the Triassic to the Early Jurassic developed several sub-basins -- Shelburne, Sable, Abenaki, South Wale, and Orpheus Graben -- and, posteriorly in a passive margin configuration, two plateaus -- Banquereau and La Have \cite{albertz2010, kettanah2013}.

During the break-up phase, which began in the Middle Triassic, the main infilling of the several interconnected sub-basins were the fluvial red bed sediments, along with volcanic rocks associated with the rifting process. In the Late Triassic, a shallow marine environment began, with the development of the Eurydice Formation, primarily formed by siliciclastic and carbonate sediments. The proper climatic configuration promoted the evaporation of the marine waters depositing salt layers, corresponding to the Argo Formation \cite{canada1997}. Subsequently, in the extent of the Early Jurassic/Late Triassic, the rifting process continued until the Break-Up Unconformity and the beginning of the proto-ocean.

Following the Jurassic succession lithostratigraphy, a shallow marine environment allowed the deposition of tidally influenced dolomites with anhydrides and siliciclastics \cite{qayyum2015}. Afterwards, the Mohican Formation sediments were deposited, which comprise muds and shales derived from a fluviomarine environment. The formation is overlaid by the Abenaki Formation, deposited in the Jurassic--Early Cretaceous during the spread of the sea floor. This formation consists of thick carbonate beds -- predominantly limestones and dolomites -- due to the configuration of a carbonate platform along the basin margin, and mudstones.


In the Late Jurassic, the Mic-Mac Formation, along with the Verril Canyon Formation and the Mohawk Formation, were deposited. These formations primarily consist of sands interbedded with shales and intercalated with carbonates, marking the initial phase of uplift and delta progradation \cite{canada1997}. In the Early Cretaceous, the Scotian Basin suffered a marine regression phase, resulting in the progradation and deposition of thick fluvio-deltaic sediments of the Mississauga Formation. Subsequently, thick shale packages with sand beds occurred due to intense deltaic sedimentation in a transgression phase, forming the Logan Canyon Formation. Still in the retrogradation phase, the Dawson Canyon Formation primarily consists of deep marine shale deposits and some limestones located across the Scotian Shelf. This sedimentation is calcareous or marly on the top, becoming shaley and silty towards its base \cite{mandal2018}.

The cessation of the deltaic sedimentation during the Cretaceous allowed the deposit of the Wyandot Formation. This deposit is the most distinctive and recognized lithologic unit
on the Scotian Shelf, consisting of chalky carbonates grading from pure chalk to marl \cite{mandal2018}. Overlying the Wyandot Formation, the Banquerau Formation consists of Tertiary marine shelf mudstones and shelf sands and conglomerates. 
This formation has a varying thickness, reaching more than 4 km beneath the continental slope, due to halokinesis \cite{canada1997}.

\begin{figure} 
    \centering
    \includegraphics[width=0.425\textwidth]{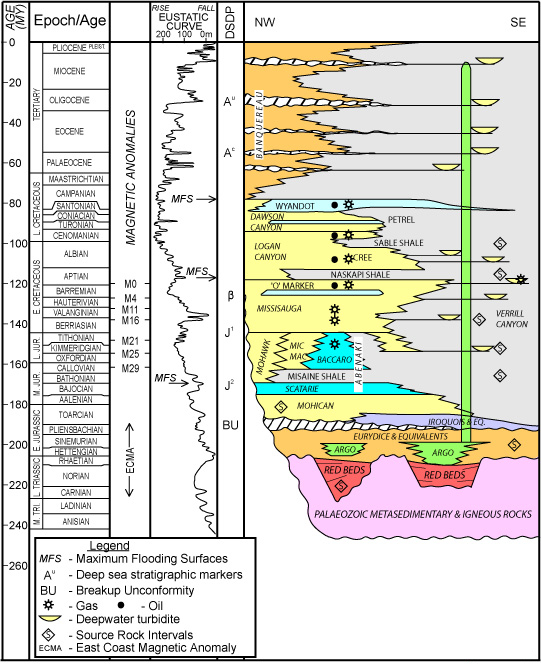}
    \caption{Chronostratigraphic chart of the Scotian Basin. (Modified from \cite{cnopb}.) (Eustatic curve from \cite{haq1988}).}
  \label{fig:chrono_chart} 
\end{figure}

The overview of the Scotian Basin lithostratigraphy and additional information about eustatic variation and possible hydrocarbon reservoirs are displayed in \autoref{fig:chrono_chart}.

\section{SEISMIC DATA}
\label{sec:seismic_data}

The seismic data used for the generation of the proposed dataset is a public 3D seismic survey called Penobscot 3D \cite{osr}, contributed by the Nova Scotia Department of Energy and the Canada Nova Scotia
Offshore Petroleum Board, and managed by dGB Earth Sciences Open Seismic Repository \cite{osrgen}. The dataset consists of 87 km\textsuperscript{2} time migrated 3D seismic data, with 601 inlines and 482 crosslines, located in offshore Nova Scotia, Canada (\autoref{fig:location}).

The seismic data has a time range of 6{,}000 ms, with 2 ms of sampling rate. The signal has a low resolution below 3{,}000 ms, approximately 5 km, with a SEG standard polarity. The acquisition parameters are 12.5 m$\times$25 m bin size (inline$\times$crossline)
with 60-fold coverage standard polarity. Along with the 3D seismic data, the repository also provides 2 wells, L-30 and B-41 -- with some markers and no geophysical logs --, pre-stack gathers, 2D seismic data, stacking velocity, and 5 interpreted horizons.



\section{SEISMIC INTERPRETATION}
\label{sec:seismic_interpretation}


The Penobscot 3D seismic dataset was imported into OpendTect \cite{opendetect} and then interpreted by two geoscientists. It is important to note that although other data are available in the repository, only the 3D seismic data were used to produce the interpretation. The interpretation was performed disregarding the horizons provided by the Open Seismic Repository \cite{osr} since they sometimes comprise more than one significant texture, what could hinder the performance of the machine learning algorithms.

\begin{figure}
    \centering
    \includegraphics[width=0.5\textwidth]{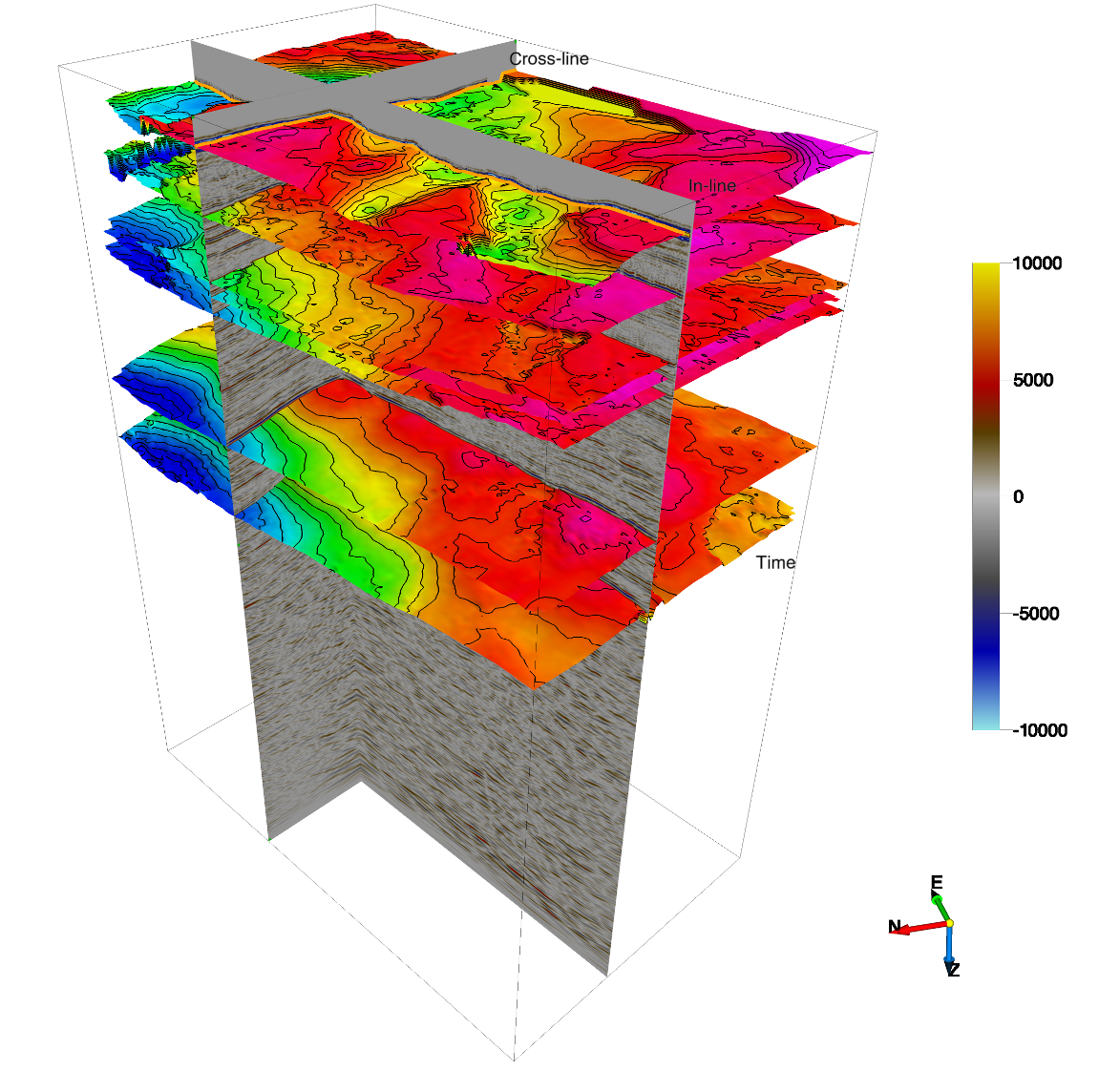}
    \caption{Seven interpreted horizons shown along with two seismic lines.}
    \label{fig:horizons} 
\end{figure}

Seven horizons were interpreted: H1, H2, H3, H4, H5, H6, and H7, numbered from the highest depth to the lowest. They divide the seismic cube into eight intervals with different pattern configurations. \autoref{fig:horizons} shows the 7 interpreted horizons along with two seismic lines. We emphasize that these horizons do not necessarily have a direct relationship with the geological settings. This means that surfaces may not correspond to the top of formations or stratal interfaces. Four faults were also interpreted to assist the horizons interpretation. For the purposes of machine/deep learning, horizon surfaces were created to separate different seismic facies intervals.

The analysis of seismic facies consists of the identification of seismic reflection parameters, based primarily on configuration patterns that indicate geological factors like lithology, stratification, depositional systems, etc. \cite{brown1980seismic}. In the following list we explain briefly the seismic facies of each of the horizon intervals based on the amplitude and continuity of reflectors.

 \begin{itemize}
   \item H1: the facies unit below H1 is characterized primarily by parallel, concordant, high-amplitude reflectors. It is also possible to identify chaotic reflectors, but that may be a consequence of the decrease of seismic frequency with depth.
   \item H2-H1: the facies unit is characterized by parallel to subparallel, continuous, high-amplitude reflectors. Parallel/subparallel configuration reflects the uniform deposition of fluvio-deltaic sediments of the Mississauga Formation.
   \item H3-H2: facies unit containing parallel to subparallel reflectors, like the previous interval, but less continuous.
   \item H4-H3: reflectors below this horizon are continuous but have low amplitude, which makes it difficult to identify them. This is expected since the sedimentary package consists of deep marine shales and limestone showing little lithological contrast.
   \item H5-H4: reflectors are predominantly subparallel and present varying amplitude.
   \item H6-H5: the package consists mostly of parallel, high-amplitude reflectors. A facies unit with chaotic seismic reflectors is also noticed and may be associated with marine slump deposits.
   \item H7-H6: the facies unit is composed of high-amplitude reflectors. Although most of the reflectors are continuous, some have diving angles and others are truncated, evidencing a high energy environment.
 \end{itemize}

\section{PENOBSCOT INTERPRETATION DATASET}
\label{sec:dataset}

\begin{figure}
\centering
\includegraphics[width=.7\linewidth]{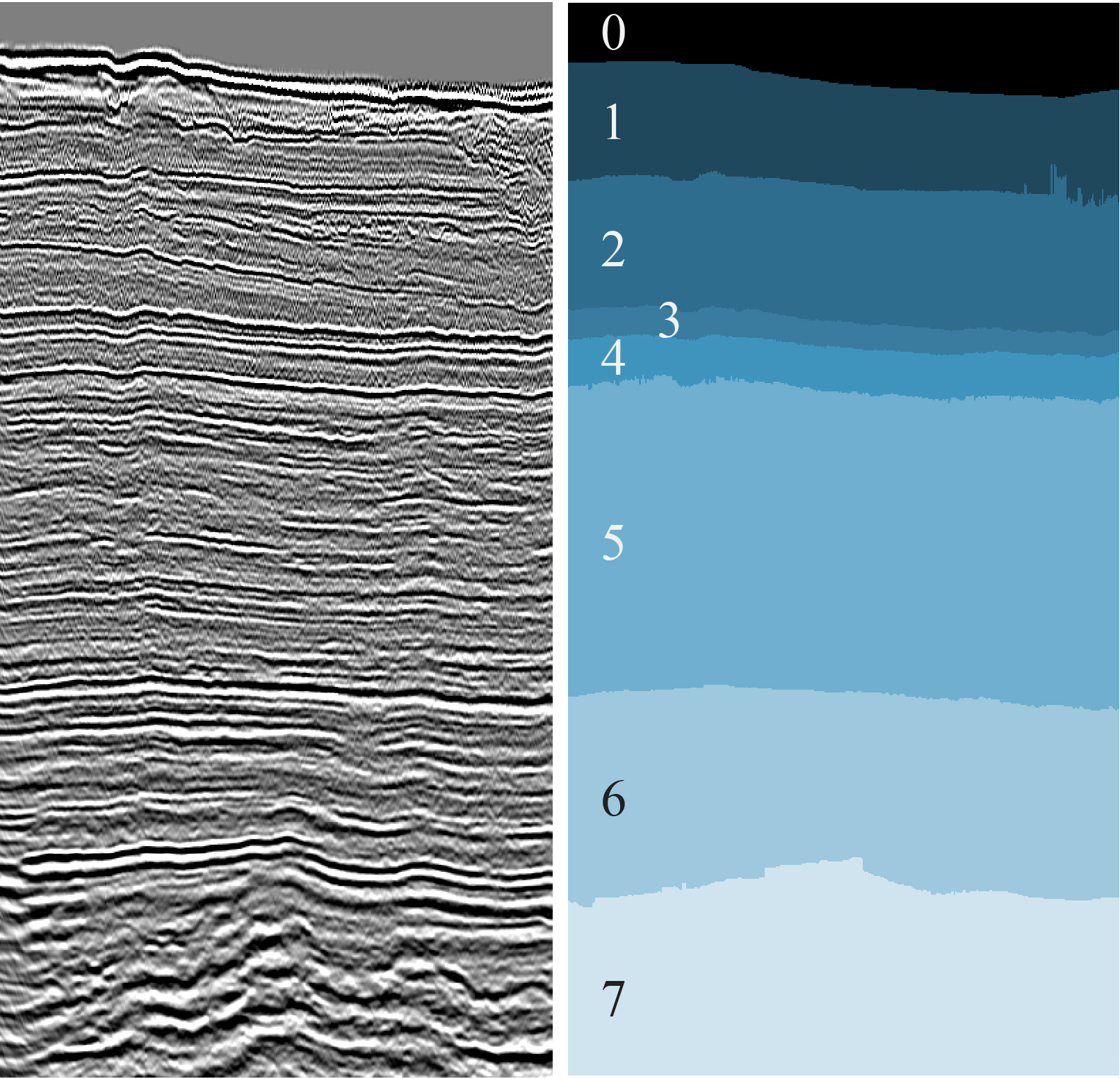}
\caption{Example of a cropped inline and its respective labels.}
\label{fig:dataset_example}
\end{figure}

The Penobscot interpretation dataset consists primarily of 7 interpreted horizons in XYZ format and 2{,}166 images (1{,}083 seismic lines in TIFF format and 1{,}083 labeled images in PNG format). To create the labeled images, we took the intersection between the horizon surfaces and each seismic line and labeled the pixels from 0 to 7, following each horizon interval. \autoref{fig:dataset_example} shows a pair of an inline image (cropped in the figure) and its respective labels.
In this paper, we present two applications: a classification and a semantic segmentation of seismic images. For the user's convenience, we provide the image tiles used in the classification task along with the dataset\footnote{The Penobscot interpretation dataset is available at: https://doi.org/10.5281/zenodo.1324463}.

To produce the classification dataset, we break the seismic images into tiles with 40$\times$40 pixels. One tile has the majority of its area belonging to only one class \cite{danets2018}. In the provided dataset, we allow 30\% of interference from other classes as discussed in \cite{chevitarese2018deep, danets2018}. The entire process of creating tiles from a seismic image comprises the following steps:

\subsubsection{split image files into training and test sets} where all tiles from a single image belong to either the training or the test set. This approach makes the training and test sets more distinct \cite{chevitarese2018deep, danets2018}.

\subsubsection{shuffle the selected images} although there are many alternatives \cite{chevitarese2018deep, danets2018} to split training and test data, we decided to simply shuffle the images, which gives a good balance between randomness and simplicity.

\subsubsection{process the images removing extreme amplitudes and re-scale values between 0 and 255} by doing this, we reduce the search space and make data smaller.

\subsubsection{generate tiles from processed images} associating them with their respective labels. Notice that labels 2 and 3 were merged during tile generation, following the discussion in \cite{danets2018, chevitarese2018deep}.

\subsubsection{balance train and test datasets} although seismic images are imbalanced regarding the areas of each seismic facies unit, machine learning methods usually rely on uniform distribution to optimize its parameters. By balancing the classes, we make the training process simpler allowing us, for example, to use well-known metrics, such as accuracy and precision.

The provided classification dataset comprises 6{,}124 crossline and 4{,}706 inline seismic tiles \textit{per class} along with their respective labels. \autoref{tab:dataset} describes the files in the dataset.

\begin{table}[]
\caption{Details of the Penobscot interpretation dataset}
\label{tab:dataset}
\centering
\begin{tabular}{@{}llrr@{}}
\toprule
File               & Format & \# Files & Total size (MB) \\ \midrule
H1-H7              & XYZ    & 7        & 87.5        \\
Seismic inlines    & TIF    & 601      & 1,700       \\
Seismic crosslines & TIF    & 481      & 1,700       \\
Labeled inlines    & PNG    & 601      & 4.9         \\
Labeled crosslines & PNG    & 481      & 3.9         \\
Seismic tiles (train) & PNG    & 75,810   & 116         \\
Seismic labels (train) & JSON    & 2        & 1.5         \\
Seismic tiles (test) & PNG    & 28,000   & 116         \\
Seismic labels (test) & JSON    & 2        & 0.5        \\ \bottomrule
\end{tabular}
\end{table}

\section{EXPERIMENTS}
\label{sec:experiments}

In this section, we present two applications that demonstrate the utility of the Penobscot dataset for seismic classification and segmentation using deep learning.

\subsection{Seismic Facies Classification}

The first application is the classification of seismic facies presented in \cite{danets2018} and \cite{chevitarese2018deep}. The authors successfully trained deep neural networks to discriminate the seismic facies present in the Penobscot dataset. They assume that one may distinguish different classes (facies) by their textural features as discussed in \cite{mattos2017assessing, chopra2006applications}.

For the presented classification task, it was necessary to break the seismic images into smaller parts (tiles), so that the majority of a tile's area belongs to only one class. Tiles are the input of the deep neural network that classifies each tile as one of the possible classes.

\begin{figure}
\centering
\includegraphics[width=.85\linewidth]{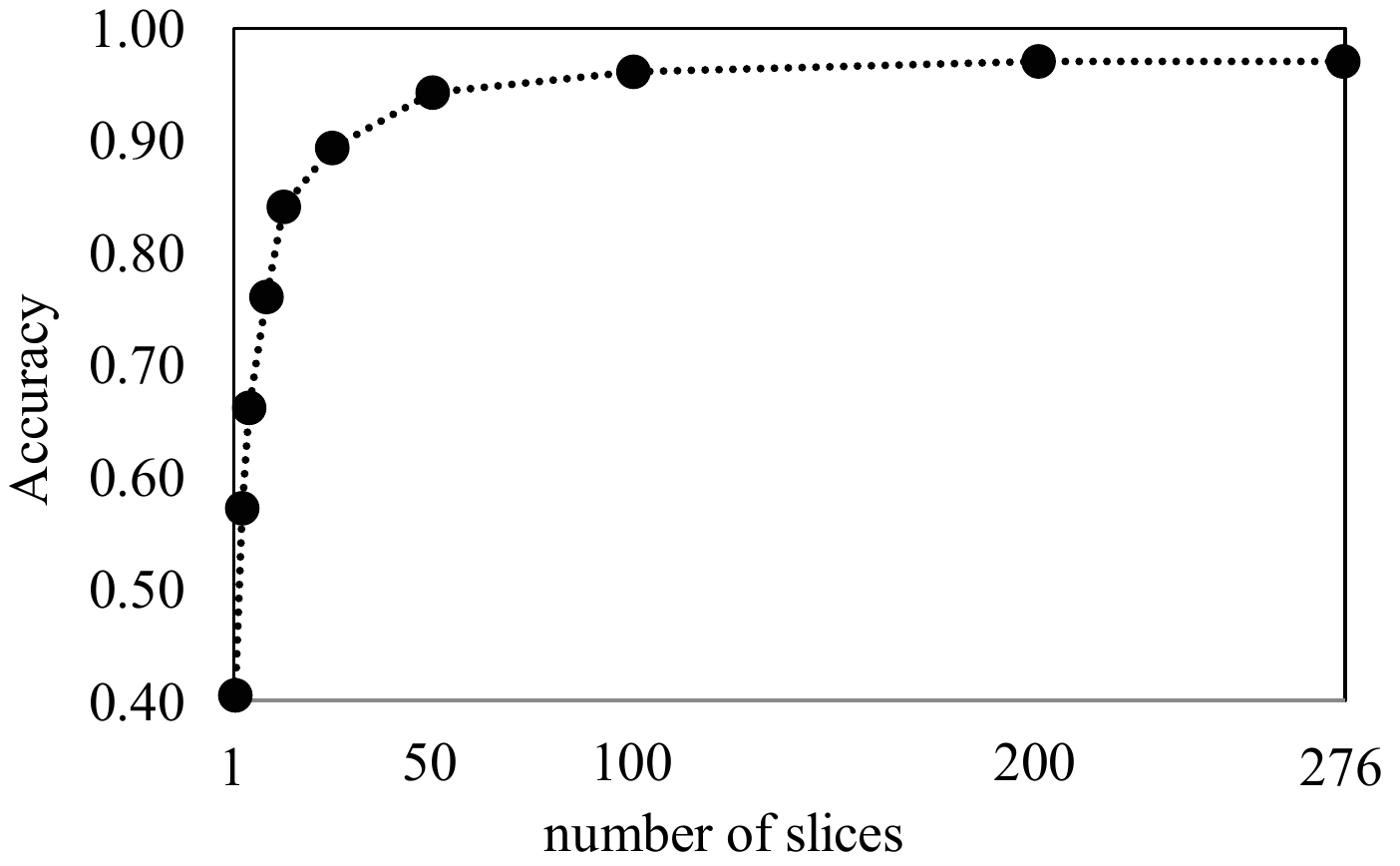}
\caption{Accuracy curve for different number of slices (1, 3, 5, 9, 13, 25, 50, 100, 200, and 276) -- plot extracted from \cite{danets2018}.}
\label{fig:accuracy}
\end{figure}

The authors tested multiple tile sizes, number of examples per class, different interference percentages among many other parameters. The results in \cite{danets2018} show that one can train a neural network in 4 minutes using 25 inline slices, and yet obtain 89\% of accuracy. Moreover, they reached up to 97\% of accuracy in 30 minutes using 276 inline slices. \autoref{fig:accuracy} shows the impact of the number of slices in the final classification accuracy.

\subsection{Seismic Facies Segmentation}

The second application in which the Penobscot interpretation dataset was used is the semantic segmentation of seismic facies. In \cite{chevitarese2018seismic}, the authors trained a deep neural network for the classification task. Then, they modified the final part of the model to produce pixel-wise predictions. Finally, they trained the resulting model using the Penobscot interpretation dataset.

\begin{figure}
\centering
\includegraphics[width=.95\linewidth]{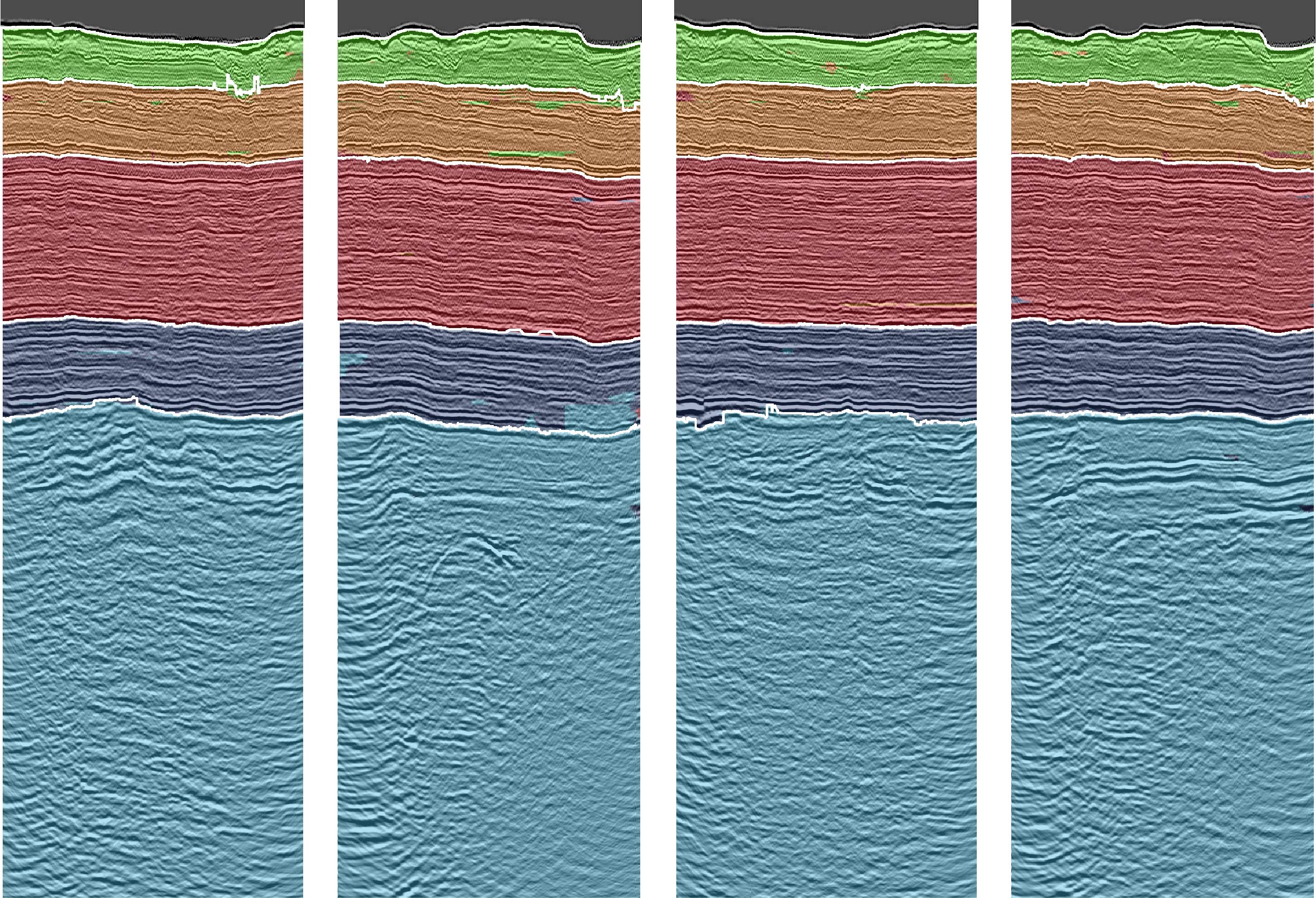}
\caption{Semantic segmentation of inlines 1200, 1520, 1090 and 1460 (left to right) from Penobscot dataset at the pixel level. In the output, each pixel receives an overlaid color representing a class. The white lines represent the seismic horizons.}
\label{fig:result}
\end{figure}

For the segmentation task, the authors also divided the input seismic images into tiles. However, the tiles are larger than the ones used for the classification task since they need to comprise more than one class. Next, they applied the network throughout the image to generate the final prediction. By doing this, they achieved more than 97\% of the intersection over union (IOU). \autoref{fig:result} shows that the model produced masks very close to the actual interpretation with very little discontinuity. Notice that the authors in \cite{chevitarese2018seismic} joined classes 2 and 3, and 4 and 5.

\section{CONCLUSIONS}
\label{sec:conclusions}

We argued in this letter that the expansion of machine learning in other fields was only possible, in part, because of the number of public datasets that have been made available in the past years. The Penobscot interpretation dataset is our contribution to foster the development of machine learning in seismic interpretation which, in our view, has gained increasing interest but still needs to build this common basis. With our dataset, we provide geoscientists, and machine learning practitioners working in the field, with +100{,}000 labeled images that can be used to develop their methods and compare their results in an easier and faster way.

In the experiments presented, the authors were able to successfully apply state-of-the-art deep learning techniques on the proposed dataset to reach high-accuracy results for seismic facies classification and segmentation. However, these are only two among many possible applications that could benefit from the dataset, such as, clustering, retrieval, and transfer learning. As future work, we intend to elaborate another dataset for other public seismic dataset called Netherlands F3. These two datasets together will provide valuable data for developing and testing machine learning methods for seismic interpretation.












\bibliographystyle{ieeetr}
\bibliography{references}

\end{document}